\documentclass[reprint,footinbib,prb,longbibliography]{revtex4-1}
\usepackage{natbib}
\usepackage{amsmath}
\usepackage{braket}
\usepackage{graphicx}
\usepackage[shortlabels]{enumitem}
\usepackage{substr}
\usepackage{verbatim}
\usepackage{import}
\usepackage{siunitx}
\usepackage{mathtools}
\usepackage{hyperref}
\usepackage{lipsum}
\usepackage{float}
\usepackage{xfrac}
\hypersetup{colorlinks=true, citecolor=blue, urlcolor=blue, linkcolor=blue}
\allowdisplaybreaks  
\begin{document}
\title{
Thermal physics in the data age --- students judge the applicability of the equipartition theorem}
\author{J.~D.~D.~Martin}
\affiliation{Department of Physics and Astronomy, University of Waterloo, Waterloo N2L 3G1 Canada}
\author{}
\affiliation{}
\date{\today}
\begin{abstract}
Providing students of introductory thermal physics with a plot of the heat capacities of many low density gases as a function of temperature allows them to look for systematic trends.  Specifically, large amounts of heat capacity data allow students to discover the equipartition theorem, but also point to its limited applicability.  Computer code to download and plot the temperature-dependent heat capacity data is provided.
\end{abstract}
\maketitle

The purpose of this paper is to point out a specific instance in the teaching of thermal physics where relatively large amounts of \emph{digital data} can help students judge the applicability of a physical concept: namely the equipartition theorem as it relates to the heat capacities of low-density gases of atoms or molecules.  Encouraging students to ``weigh the evidence'' helps them practice a skill of working scientists, who often have to decide if a paper's conclusions are warranted by its data.

The organization of this paper is as follows: first the equipartition theorem is briefly reviewed.  Then the common textbook example of its application to H$_2$ gas is presented.  But the case of H$_2$ is not representative, as is shown by plotting the heat capacities of numerous molecules as a function of temperature.  This limited applicability of the equipartition theorem is discussed in the context of introductory thermal physics courses.  Finally, some insight is gained by considering a simple model for the temperature-dependent heat capacity of gaseous CO$_2$, which is suitable for study in statistical mechanics courses.

The equipartition theorem is a result of \emph{classical} statistical mechanics.  For a precise statement and thorough derivation see, for example, Ref.~\onlinecite{isbn:9781577666127}. Roughly speaking, the theorem says that each degree of freedom of a system that contributes a quadratic term to the energy adds $k_BT/2$ to the total average energy of the system, where $k_B$ is Boltzmann's constant and $T$ is the absolute temperature. Following Ref.~\onlinecite{isbn:978-0-201-38027-9}, I denote the number of these contributions as $f$ (per atom or molecule).  In that way, a system of $N$ non-interacting atoms or molecules --- each with $f$ degrees of freedom contributing quadratic terms to the energy --- has a total energy of $U = f Nk_BT/2$, and thus a temperature-independent constant volume heat capacity of $C_V = f N k_B/2$.

In this paper, discussion will be restricted to non-interacting atoms and molecules; i.e., low-density gases following the ideal gas law.  Nevertheless, the equipartition theorem is of broader applicability, most notably here to the heat capacities of single-element solids, as expressed by the Dulong--Petit law, where $f=6$; see, for example, Ref.~\onlinecite{shortdoi:c3ncrc}.  It also played an important --- but sometimes misrepresented\cite{shortdoi:gfkbjf} --- role in the history of understanding thermal electromagnetic radiation; i.e., the Rayleigh–Jeans ultraviolet catastrophe.

The cleanest application of the equipartition theorem is to monatomic ideal gases.  Under the assumption that we can ignore their internal structure, there are three quadratic terms in the energy per atom: $U/N = p_x^2/(2m) + p_y^2/(2m) + p_z^2/(2m)$, where $p_i$ are the momenta in three orthogonal directions and $m$ is the mass of each atom. So $f=3$ and thus $C_V=3 Nk_B/2$ for monatomic ideal gases, if the equipartition theorem applies.

However, the equipartition theorem assumes classical --- not quantum-mechanical --- motion.  In most cases, the motion of monatomic gases can be understood classically, and thus the theorem works quite well.\footnote{Gottfried and Yan note (\emph{Quantum Mechanics: Fundamentals}, Springer, NY, 2003) that with the benefit of hindsight the lack of an \emph{electronic} contribution to atomic heat capacities (at low temperatures) points to the deficiencies of classical mechanics.}  But once we consider molecules, we are immediately confronted --- as were the founders of quantum mechanics --- by the limited applicability of the equipartition theorem due to the quantization of molecular rotation and vibration.\cite{shortdoi:b29vxg}

\begin{figure}[!b]
\includegraphics{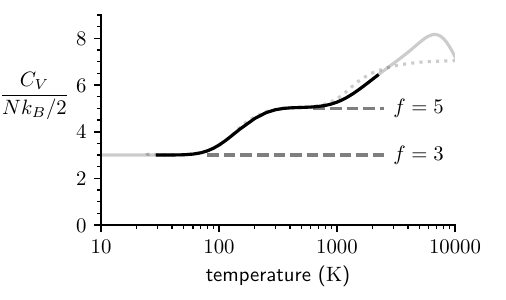}
\caption{\label{fg:h2_heat_capacity}
The scaled heat capacity of low-density gaseous H$_2$ as a function of temperature,\cite{shortdoi:br98wj} assuming a constant 3:1 mixture of ortho- and para-hydrogen (so-called \emph{normal hydrogen}\cite{shortdoi:bjdpb2}). This plot is modelled after Fig.~1.13 of Schroeder \cite{isbn:978-0-201-38027-9} (dark line) but is extended over a larger temperature range (by the light solid line) using the results of Ref.~\onlinecite{shortdoi:br98wj}.  The light dashed line is as shown in Fig.~20.6 from Serway {\it et al.},\cite{isbn:9781337553292} who do not provide details on its origin.}
\end{figure}

\begin{figure*}
\centering
\includegraphics{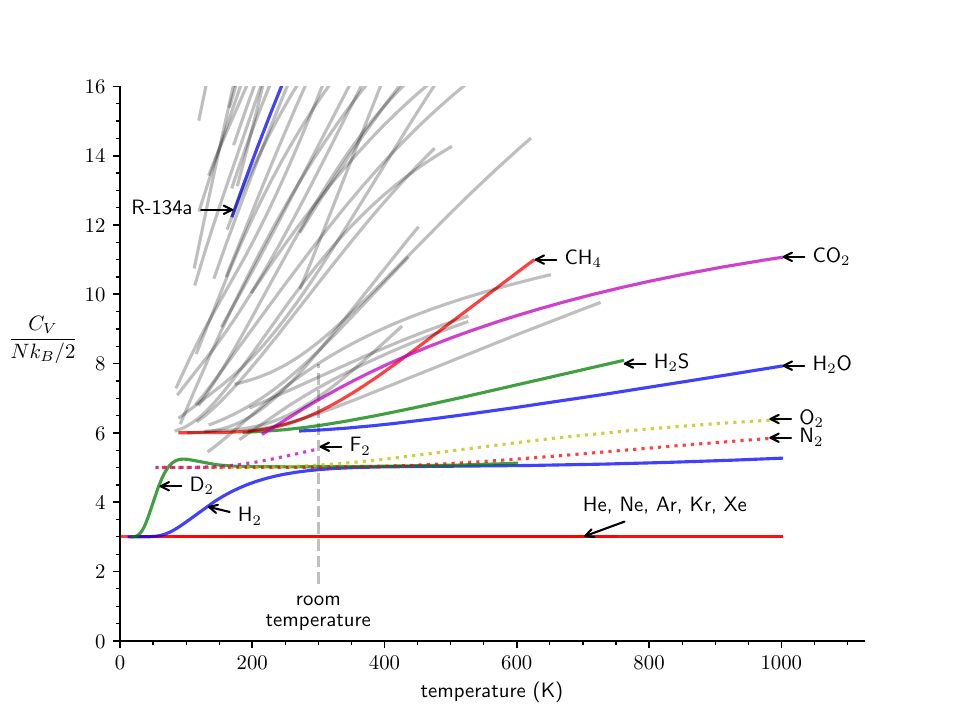}
\caption{~(color online)
\label{fg:molecular_heat_capacities}
The scaled heat capacities of low density gases as a function of temperature, as obtained from the NIST Chemistry Webbook.\cite{shortdoi:ghxvrw:modified}
All monatomic and diatomic species are labeled.  All unlabeled curves correspond to triatomic or larger species.  Many of these molecules are refrigerants, reflecting the technological importance of their thermal properties; see for example, Ref.~\onlinecite{shortdoi:f9rbgx}.  For example, the {\bf R-134a} label on the plot refers to 1,1,1,2-tetrafluoroethane (CH$_2$FCF$_3$), once commonly used in automobile air-conditioners.  To give an objective view of the validity of equipartition theorem, all molecules in the database with $C_V/(Nk_B/2) < 16$ are shown, except to prevent crowding, I have omitted carbon monoxide, deuterium oxide, and the ortho- and para- variants of hydrogen and deuterium (the normal mixtures are plotted).  (See the supplemental material\cite{supplemental} for the Python program to download the data and create this plot.)
}
\end{figure*}

An illustrative example of the applicability of the equipartition theorem to low density H$_2$ gas appears in a popular and well-respected textbook on thermal physics by Schroeder \cite{isbn:978-0-201-38027-9} --- see Fig.~\ref{fg:h2_heat_capacity}.\cite{[{Schroeder\cite{isbn:978-0-201-38027-9} gives an early and significant role to the equipartition theorem compared to other textbooks at a similar level; for example, instead of giving the derivation of the formulae for adiabatic, quasistatic compression of an ideal gas in terms of $\gamma \coloneqq C_P/C_V$, Schroeder uses $f$, compared to the more conventional usage of $\gamma$; e.g.,\phantom{x}}][]isbn:978-0-19-956209-1}
Figure \ref{fg:h2_heat_capacity} shows that below about \SI{100}{K}, H$_2$ behaves like a monatomic gas, with only the translational degrees of freedom active.  At these low temperatures, molecular rotation is ``frozen'' out due to quantization of the rotational energy level structure.   But from about \SI{300}{K} to \SI{1000}{K} --- once rotation has been activated --- the equipartition theorem appears to be applicable again, but with $f=5$, the two additional quadratic degrees of freedom corresponding to rotation.\cite{[][{. As Gearhart notes, an explanation of why the rotation of diatomic molecules only contributes two and not three extra degrees of freedom to $f$ --- seemingly avoiding rotation about the axis connecting the atoms --- has been a stumbling block for many textbook authors.  A thorough explanation requires a knowledge of quantum mechanics beyond what would be typical of students in an introductory thermal physics course.}]shortdoi:fv7mr9}  At about $\SI{1000}{K}$, the heat capacity begins to increase again due to the activation of vibrational motion.

That a theorem based on \emph{classical} mechanics can be applied to a quantum mechanical system \emph{over limited temperature ranges} is not surprising.  After all, although the center-of-mass motion of atoms and molecules should in principle be described using quantum mechanics, as far as heat capacities go, the classical picture is normally sufficient; i.e., $C_V = 3 Nk_B /2$ if there are no internal excitations of the atoms/molecules.   Reif\cite{isbn:9781577666127} indicates that the criteria for approximate validity of the equipartition theorem is that the spacing of energy levels around the mean energy be small compared to $k_BT$.  For typical confining volumes, the translational energy levels may be considered so dense as to be continuous, and thus the equipartition theorem is virtually always applicable to the translational degrees of freedom of atoms or molecules.

Since the rotational energy level spacings in molecules are larger than those for translation but typically smaller than those for vibration, we might expect a temperature range where the equipartition theorem gives $C_V = (f_t + f_r)Nk_B/2$ with the contribution due to translation of $f_t=3$ and $f_r = 2$ or $f_r=3$ due to rotation, depending on whether the molecular geometry is linear ($f_r = 2$; e.g, H$_2$, CO$_2$) or non-linear ($f_r=3$; e.g., H$_2$O, CH$_4$).\cite{isbn:9781891389054} As noted above, H$_2$ does approximately exhibit a temperature-independent heat capacity
 $C_V \approx (f_t + f_r)Nk_B/2$ with $f_t=3$ and $f_r=2$ from $\approx \SI{300}{K}$ to $\approx \SI{1000}{K}$.

At higher temperatures, we expect vibration to become important.  The simplest model for molecular vibration involves a ``linearization'' of forces and subsequent decomposition of internal motions into normal modes, each acting as an independent harmonic oscillator.\cite{isbn:9780894642692}  Since harmonic oscillators have two degrees of freedom (momentum and position) contributing quadratic energy terms to the total energy, we might expect from the equipartition theorem that there will be a temperature range for which $C_V \approx (f_t + f_r + f_v) Nk_B/2$ with the contribution due to vibration $f_v$ being equal to twice the number of vibrational modes that a molecule has.  For instance, for H$_2$, $f_v=2$, since a diatomic molecule only has one vibrational mode (corresponding to the relative motion of the two nuclei). However, as evident from Fig.~\ref{fg:h2_heat_capacity}, there is no extended temperature range over which the heat capacity for H$_2$ is given by $C_V = f Nk_B/2$, where $f=7$ (as expected from $f=f_t+f_r+f_v$ with $f_t=3$, $f_r=2$, and $f_v=2$).  As illustrated in Fig.~\ref{fg:h2_heat_capacity} and discussed later, some textbooks\cite{isbn:9781337553292} present plots suggesting that H$_2$ \emph{does} exhibit a heat capacity plateau corresponding to $f=7$, at odds with the research literature.\cite{shortdoi:br98wj}

A phenomenologically based discussion of heat capacities is common in first-year general university physics courses and towards the start of upper-year thermal physics courses (before a more in-depth treatment of their statistical mechanical origin).\cite{isbn:978-0-201-38027-9}  Some commonly used textbooks\cite{isbn:9781337553292, isbn:9781119460138} show a version of Fig.~\ref{fg:h2_heat_capacity} and briefly discuss the equipartition theorem, without deriving it.  In this context, it may be useful to show students many more examples of temperature-dependent heat capacities to help them gauge how general the behavior of Fig.~\ref{fg:h2_heat_capacity} is; i.e., do other molecules have temperature ranges with temperature-independent heat capacities, given by $C_V = f Nk_B/2$, with $f$ a positive integer, depending on the temperature range?

For this purpose, Figure \ref{fg:molecular_heat_capacities} shows the temperature dependence of large number of heat capacites, downloaded from the National Institute of Standards and Technology (NIST) Chemistry Webbook.\cite{shortdoi:ghxvrw:modified}  Rather than choosing molecules that illustrate the equipartition theorem, all of the available data\footnote{I follow NIST's usage of the term ``data''; the heat capacities and other tabulated quantities in Ref.~\onlinecite{shortdoi:ghxvrw:modified} are based on mixtures of calculations and empirical data; i.e., some heat capacities are computed using spectroscopically determined energy levels.} has been plotted, with just a few species omitted for clarity (which can be added by slightly modifying the plotting code provided in the supplemental material\cite{supplemental}).

Figure \ref{fg:molecular_heat_capacities} shows that in general the equipartition theorem is less applicable than one might think from the example of H$_2$ over the temperature range shown by the dark line in Fig.~\ref{fg:h2_heat_capacity}.  For the larger polyatomics --- unlabeled in Fig.~\ref{fg:molecular_heat_capacities} --- it does not apply at all, at least in the temperature range shown.  There are no clearly isolated steps in $C_V$ with temperature because of the variety of low and high vibrational mode frequencies,\cite{[][{. This book contains illustrative examples of the temperature dependence of gas phase heat capacities.}]isbn:978-0486842745} as will be seen below by examining the case CO$_2$ in more detail.

In an introductory thermal physics course, one could introduce Fig.~\ref{fg:molecular_heat_capacities} in the following ahistorical manner: after derivation of the relationship between the average kinetic energy of a gas molecule and its temperature: $U/N = 3 k_BT /2$ (see pages 10 -- 11 of Ref.~\onlinecite{isbn:978-0-201-38027-9}), Fig.~\ref{fg:molecular_heat_capacities} could be examined with the (erroneous) expectation that $C_V = 3 Nk_B/2$, independent of temperature.  The difference between the atomic and molecular cases will naturally lead to
discussion of the additional degrees of freedom of molecules --- beyond translation --- that result in larger heat capacities.  Also
deviations from $C_V = 3 Nk_B/2$ due to electronic excitations of atoms may be anticipated, although they are not observed for the atoms and temperature ranges of Fig.~\ref{fg:molecular_heat_capacities}.

From Fig.~\ref{fg:molecular_heat_capacities} students will be able to identify that --- besides $C_V = 3 Nk_B /2$ --- there is something special about certain other values of the heat capacity. Specifically, they may observe:
\begin{enumerate}[(1),itemsep=-2pt]
\item the transition to $C_V = 5 Nk_B/2$ for hydrogen, the propensity to this $C_V$ for other diatomics (within limited temperature ranges), and also that
\item the labeled triatomics (except for CO$_2$) show a propensity towards $C_V = 6 Nk_B /2$ at low temperatures; i.e, $f=6$, exhibiting an additional rotational degree of freedom beyond the diatomic case.
\end{enumerate}

\begin{figure}[b]
\centering
\includegraphics{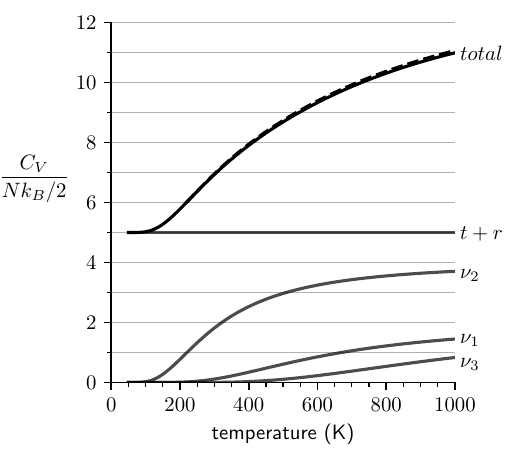}
\caption{\label{fg:co2}
The scaled heat capacity of low-density gaseous CO$_2$ ($total$) together with the individual contributions of translation and rotation ($t+r$), and the three vibrational modes ($\nu_1$, $\nu_2$, and $\nu_3$) as calculated in a simplified model (see text).  The dashed line is the data from the NIST Chemistry Webbook.\cite{shortdoi:ghxvrw:modified}}
\end{figure}

\begin{figure}[t]
\centering
\includegraphics{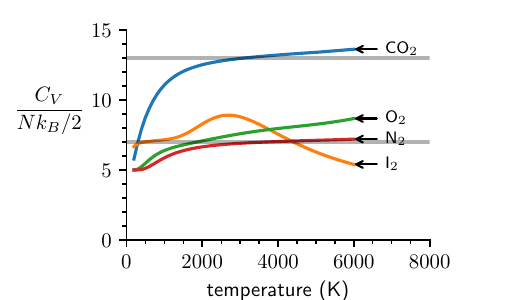}
\caption{~(color online) \label{fg:anharm}
The scaled heat capacities of some low density gases as obtained from the NIST-JANAF database.\protect\cite{[][{. Quantities from this database may be accessed using the Python package available at \protect\url{https://thermochem.readthedocs.io};
Last accessed on 2023-03-19.}]isbn:9781563968310}}
\end{figure}

With this background, the equipartition theorem can be mentioned, with its limited applicability obvious to students from Fig.~\ref{fg:molecular_heat_capacities}.  Further discussion may then be deferred until statistical mechanics is studied.  Until then, the equipartition theorem is just a handy way to remember commonly used heat capacities: $C_V = f Nk_B/2$, where $f=3$ for atomic species without electronic excitation, and $f \approx 5$ for N$_2$ and O$_2$ near room temperature, which is particularly useful as so many problems in introductory thermal physics involve the heat capacity of air.  (Figure \ref{fg:molecular_heat_capacities} also gives students a sense of the temperature range over which $f\approx 5$ is a good approximation for air.)  I have found that, by following this approach, students' misconceptions regarding the applicability of equipartition theorem have largely been eliminated.\footnote{When following Ref.~\onlinecite{isbn:978-0-201-38027-9} quite closely --- giving prominence to Fig.~\ref{fg:h2_heat_capacity} --- I would frequently observe the appearance of $C_V = fNk_B/2$ as a \emph{general} expression for heat capacity in students' work, including situations where the equipartition theorem was not applicable.  This misuse was possibly due to the ``theorem'' terminology, suggesting a good opportunity to discuss the role of precise mathematical results in physics: their premises are often only approximately satisfied in the physical world.
}

Once students begin their study of statistical mechanics, they may benefit from a discussion of the role of heat capacities and the equipartition theorem in the historical development of quantum mechanics.\cite{shortdoi:b29vxg, shortdoi:bv6tj7}

Furthermore, aspects of Fig.~\ref{fg:molecular_heat_capacities} are suitable for study in introductory statistical mechanics.  For example, to gain some insight into the limited applicability of the equipartition theorem we may consider a simple model for the temperature dependence of the heat capacity of CO$_2$.\cite{isbn:9780894642692, isbn:9781891389054}

The CO$_2$ molecule has three distinct vibrational mode frequencies, commonly denoted by
$\nu_1$ (symmetric stretch),
$\nu_2$ (bending), and \
$\nu_3$ (antisymmetric stretch).\cite{isbn:9780199382576}
Defining corresponding characteristic temperatures $\Theta_i \coloneqq h \nu_i / k_B$, where $h$ is Planck's constant, we have
$\Theta_1 \approx \SI{2000}{K}$
$\Theta_2 \approx \SI{960}{K}$, and
$\Theta_3 \approx \SI{3380}{K}$.\footnote{The mode frequencies are taken from the fundamental bands listed in Table 56 of Herzberg \cite{isbn:9780894642692} as he recommends on page 504.}
If we assume that each mode acts as an independent harmonic oscillator, then each makes an additive contribution to the heat capacity, given by a standard result often derived in the context of the Einstein solid model (e.g., Eq.~7.7.5 of Ref.~\onlinecite{isbn:9781577666127}):
\begin{equation}\label{eq:c_einstein}
\frac{C_{V,i}}{\frac{1}{2} Nk_B} = 2 g_i
\left(
\frac{\Theta_i}{T}
\right)^2
\frac{ e^{\Theta_i/T} }{(e^{\Theta_i/T} - 1)^2},
\end{equation}
where the $g_i$'s are the degeneracies of each mode.  The bending mode of CO$_2$ ($\nu_2$) is doubly-degenerate, so $g_2 = 2$; otherwise $g_1=g_3=1$.

Figure \ref{fg:co2} shows the individual mode contributions to the heat capacity of CO$_2$ as computed using this simple model.  Over the temperature range plotted, the equipartition theorem applies to the rotational motion, so that the total heat capacity is:
\begin{equation}
C_V = (f_t + f_r) \: Nk_B / 2 \: + \sum_{i=1,2,3} C_{V,i}
\end{equation}
with $f_t=3$ and $f_r=2$ ($f_r=2$ instead of $3$ since CO$_2$ is a linear molecule\cite{isbn:9781891389054}), and the $C_{V,i}$'s given by Eq.~\ref{eq:c_einstein}.

In the high-temperature limit, Eq.~\ref{eq:c_einstein} gives $C_{V,i} = g_i N k_B$ for $g_i$ harmonic oscillators; i.e., $f=2$ for each oscillator, as expected from the equipartition theorem.\cite{isbn:9781577666127}  Thus we expect a jump of $\Delta f = 2 g_i$ as each mode is activated with increasing temperature.  Since the $\nu_2$ bending mode has the lowest frequency, we expect it to be activated at a lower temperature than the $\nu_1$ and $\nu_3$ modes.  However, Fig.~\ref{fg:co2} shows that the activation of the $\nu_1$ and $\nu_3$ modes obscure observation of the $\nu_2$ plateau.

Furthermore, the low frequency of the $\nu_2$ bending mode limits the range over which the $f=5$ (translation and rotation) plateau may be seen.
This behavior is typical of the larger molecules in Fig.~\ref{fg:molecular_heat_capacities} which usually have low frequency bending and hindered internal rotation modes.\footnote{``Hindered internal rotation'' refers to the rotation of groups of atoms \emph{within} a molecule (e.g,. a methyl group), where the rotation encounters potential barriers.  See, for example, Ref.~\onlinecite{isbn:9780894642692}.}
\cite{isbn:9780894642692}

In contrast, the limited number of small polyatomics in Fig.~\ref{fg:molecular_heat_capacities} that show a propensity to $f=6$ around room temperature (H$_2$S, H$_2$O, CH$_4$) do not have low-frequency modes (also these molecules all have a non-linear geometry and thus rotation contributes $f_r=3$ in contrast to $f_r=2$ for CO$_2$ which is linear\cite{isbn:9781891389054}).  For example the lowest vibrational mode frequency of H$_2$O has $\Theta_2 \approx \SI{2300}{K}$ in comparison to the lowest for CO$_2$: $\Theta_2 \approx \SI{960}{K}$.  The delayed activation of vibration in H$_2$O facilitates the observation of an $f=6$ plateau.

Now let us consider the high-temperature limit of the simple model for the heat capacity of CO$_2$,  decomposing the contributions due to translation $t$, rotation $r$, and vibration $v$ as: $f=f_t + f_r + f_v$ in $C_V = f Nk_B/2$.  From the $T \rightarrow \infty$ limit of Eq.~\ref{eq:c_einstein} together with the number of modes and their degeneracies, $f_v = 8$.  Thus we expect a heat capacity corresponding to $f = 13$ at high temperatures (from $f_t=3$, $f_r=2$, and $f_v=8$).  But as the higher temperature heat capacity data for CO$_2$ in Fig.~\ref{fg:anharm} shows, this limit is approached and then exceeded as $T$ increases.  It is natural to expect this breakdown of the harmonic oscillator model at such high temperatures, for, at the very least, all molecules dissociate and thus do not have an unbounded vibrational energy level structure (for CO$_2$ from Ref.~\onlinecite{shortdoi:f7m3cc}: $D/k_B \approx \SI{6e4}{K}$ where $D$ is the energy required to break one of the C-O bonds).

Returning to the case of H$_2$, the situation is similar to that of CO$_2$: there is no $f=7$ plateau ($f_t=3$, $f_r=2$, $f_v=2$) observed in $C_V$ (the light solid line of Fig.~\ref{fg:h2_heat_capacity}), contrary to the misleading depiction in Ref.~\onlinecite{isbn:9781337553292} (the light dotted line of Fig.~\ref{fg:h2_heat_capacity}).  As illustrated in Fig.~\ref{fg:anharm}, the lack of a $f=7$ plateau is quite typical of diatomics.  Its absence may be accounted for by a simple model for anharmonicity in the potential describing the interaction of the two atoms.\cite{shortdoi:j3dd} This model may serve as a useful computational exercise for students of statistical physics.   Accounting for anharmonicity in polyatomics is significantly more challenging.\cite{shortdoi:j3dd}

In summary: giving students of introductory thermal physics the opportunity to examine the heat capacities of a variety of molecules allows them to develop a better appreciation of the applicability of the equipartition theorem than is normally the case when the heat capacity of H$_2$ is presented over a limited temperature range.\cite{isbn:9781337553292, isbn:9781119460138, isbn:978-0-201-38027-9} Students of statistical mechanics may explore the deviations from the equipartition theorem in more detail, as illustrated here using a simple model for the heat capacity of CO$_2$.

Finally, it is noted that there is a wealth of digital thermodynamic data available for teaching thermal physics.  A particularly useful resource is the CoolProp library.\cite{[][{;  see \url{http://coolprop.org}, Last accessed on 2023-01-29}]shortdoi:f5srdv}  For example, using the Python interface to CoolProp, students may explore non-ideal fluid properties, such as deviations from the ideal gas law, and predict the thermodynamic efficiencies of refrigerators using different fluids; e.g., Section 4.4 of Ref.~\onlinecite{isbn:978-0-201-38027-9}. The examination of digital thermodynamic data --- that has only recently become widely available --- gives students an active role in judging the applicability of what they are learning.

The author thanks J.~C.~T.~Martin, N.~Fladd, U.~Nandivada, and M.~Robbins for comments and V.~Koottala for checking the portability of the Python code.  This work was supported by NSERC (Canada).

\bibliography{005_references+extras,006_references+references}

\end{document}